\documentclass[journal]{IEEEtran}

\ifx\pdfoutput\undefined
\usepackage{graphicx}
\else
\usepackage[pdftex]{graphicx}
\fi

\begin{document}
\title{Quantitative Theory of Nanowire and Nanotube Antenna Performance}
\author{P.~J.~Burke$^*$,~\IEEEmembership{Member,~IEEE,}
        S. Li,~\IEEEmembership{Member,~IEEE,}
        and~Z.~Yu,~\IEEEmembership{Member,~IEEE}% <-this % stops a space
\thanks{The authors are with the Integrated Nanosystems Research Facility, Department of Electrical Engineering and Computer Science, University of California, Irvine, CA, 92697-2625. This work was supported by the ONR, NSF, ARO, and DARPA.}% <-this % stops a space
\thanks{$^*$Corresponding author: pburke@uci.edu}}

%\markboth{preprint draft \today}{preprint draft \today}

%\date{Today}
\maketitle

\begin{abstract}
We present quantitative predictions of the performance of nanotubes and nanowires as antennas, including the radiation resistance, the input reactance and resistance, and antenna efficiency, as a function of frequency and nanotube length. Particular attention is paid to the quantum capacitance and kinetic inductance. In so doing, we also develop a circuit model for a transmission line made of two parallel nanotubes, which has applications for nano-interconnect technology.
\end{abstract}


\IEEEpeerreviewmaketitle

\section{Introduction}

\PARstart{W}e recently demonstrated the operation of active nanotube devices at microwave (GHz) frequencies\cite{2004:753-756}. However, the electrical properties of nanotubes as passive high frequency components such as interconnects, mixers, detectors, and antennas are currently not well understood. In this work, we study theoretically the interaction of one dimensional electronic systems with microwave radiation, leading to a quantitative theory of nanowire and nanotube antenna performance.

In our previous modeling work
\cite{2003:55-58,2002:129-144}, 
we briefly considered nanotubes as antennas but did not quantitatively assess their performance potential. Recently, we have been able to synthesize and electrically contact
single-walled carbon nanotubes (SWNTs) up to $\sim$~1~cm in length
\cite{2004:00,2004:01}. These tubes are comparable in length to the  wavelength of microwaves in free space. This motivates our study of the interaction of microwaves with nanotubes, and the exploration of their properties as antennas.

Nanotubes grown in our lab have conductivities several times larger than copper
\cite{2004:00,2004:01}, 
but the diameter is small, so the resistance is high. Thus, current nanotube growth technology allows for very lossy antennas. In spite of heavy losses, these may allow a wireless, non-lithographic connection between nanoelectronic devices and the macroscopic world. If lower resistance nanotubes can be grown, we predict the antenna properties to be dramatically different from conventional thin-wire antennas.

\subsection{Limits of applicability}

The geometry we consider is that of a thin-wire, center fed antenna where the wire is made of a single walled metallic carbon nanotube. This is the first step to a general theory of nano-antennas. Our calculations should also apply to semiconductor nanowire antennas in the quantum mechanical 1d limit, and also to multi-walled nanotube (MWNT) antennas if suitably generalized. Our theory applies only in the quantum mechanical 1d limit, where only one sub-band is occupied by the electrons.

Therefore, this work does {\it not} apply to metallic ``nanowires'' (which are usually not in the 1d quantum limit), nor to semiconducting nanowires with more than one occupied sub-band. A possible future project would be to determine the crossover from nano-antenna to thin-wire antenna behavior. Some work in this intermediate regime has recently begun\cite{2002:65-74,2002:339-339,2003:735-745}, and we discuss this crossover more extensively below.

Our work should apply in the microwave, sub-mm, and THz spectrum, but not in the optical or IR spectrum. In the latter, the photon energy is sufficiently large that electronic excitations must be taken into account.

\subsection{Outline}

This article is divided up as follows: First, we discuss state-of-the art in nanotube synthesis, paying particular attention to nanotubes with length of order the wavelength of microwaves, i.e. cm\cite{2004:00,2004:01}. Then, we discuss possible applications of nanotube antennas. Third, we present a circuit model for a two-nanotube transmission line, a necessary pre-step for the following sections. Fourth, based on this circuit model, we calculate the spatial current distribution for a nanotube antenna. Once this current distribution is known, we treat each infinitesimal element of current as a radiator and add up (integrate) their contributions 
to the electric field to determine the total far-field electric field, hence radiated power. We do this first in the no ohmic loss case, and then in the low ohmic loss case, and finally in the high ohmic loss case.

Where possible we provide executive summary type conclusions of our calculations for performance predictions to researchers interested in building and measuring the performance of nanotube antennas. In what follows, we use the same terminology and symbol definition as our prior papers\cite{2003:55-58,2002:129-144}.

\section{Nanotube growth state-of-the-art}

In our lab, we have grown the longest electrically contacted SWNTs, with lengths up to 0.4 cm in length. Our measurements indicate that the resistance per unit length is around $10~k\Omega/\mu m$. When scaled by the diameter of 1.5~nm, this gives rise to a conductivity of $10^9~S/m$ (resistivity of $0.1~\mu\Omega-cm$), which is 10 times larger than copper. A similar conductivity was measured on 300~$\mu$m long SWNTs by one other group\cite{2004:35-39}. Two other groups have been able to synthesize 600~$\mu$m\cite{2002:703-708} and several mm to 1.1 cm\cite{2003:13251-13254,2003:5636-5637,2003:1651-1655,2004:1025-1028,2004:1077-1079} SWNTs. Therefore, the synthesis of long SWNTs is possible in several laboratory settings. 

For the resistance per unit length that we measured, as we show below, this would correspond to a very heavily damped antenna, with significant ohmic losses. However, the mechanism for the scattering in long SWNTs is still not well-studied. With sufficient effort it may be possible to lower the resistance per length by improving the synthesis technique. Therefore, the prospects for low-loss antennas, while a long-term possibility, are not entirely unreasonable.

\section{Applications of nanotube antennas}

\subsection{A solution to the nano-interconnect problem}

Progress to date on nanoelectronics has been significant. Essentially all devices needed to make the equivalent of a modern digital or analog circuit out of nanotubes and/or nanowires have been demonstrated in prototype experiments, and elementary logic circuits have been demonstrated\cite{2001:1317-1320,2001:453-456,2002:929-932,2001:1313-1317}. 

However, one of the most important unsolved problems in nanotechnology is how to make electrical contact from nanoelectronic devices to the macroscopic world, without giving up on the potential circuit density achievable with nanoelectronics. 

All of the nanotube and nanowire devices developed to date have been contacted by lithographically fabricated electrodes. A canonical research theme is to fabricate a nanodevice, contact it with electrodes fabricated with electron-beam lithography, then publish a paper reporting the electrical properties. This is not a scalable technique for massively parallel processing, {\it integrated nanosystems}. The potential high-density circuitry possible with nanowires and nanotubes will not be realized if each nanowire and nanotube is contacted lithographically.

\begin{figure}
\centering
\includegraphics[width=3.5in]{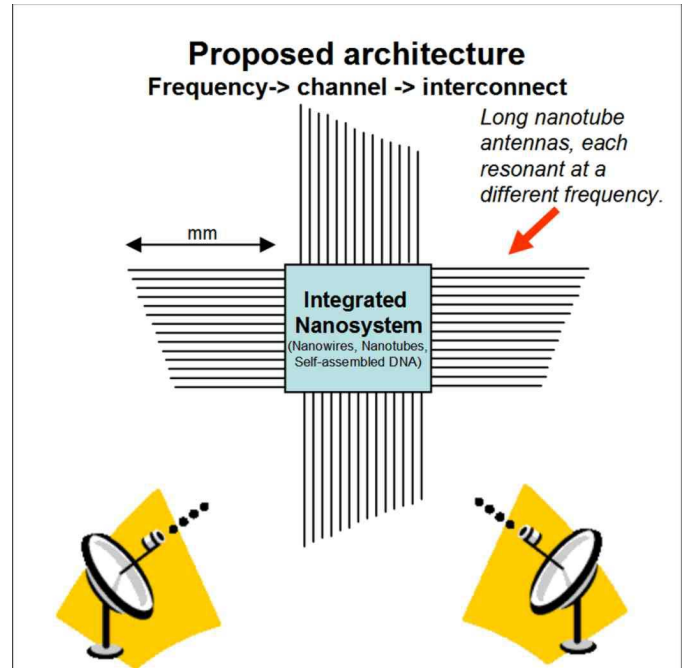}
\caption{Possible scheme for wireless interconnection to integrated nanosystems.}
\label{fig:INS}
\end{figure}

One potential solution to this problem is to use wireless interconnects, which can be densely packed. If each interconnect is connected to a nanotube of a different length (hence different resonant frequency), then the problem of multiplexing input/output signals can be translated from the spatial domain to the frequency domain, hence relaxing the need for high resolution (high cost) lithography for interconnects. This is in contrast to previous approaches which, ultimately, rely on lithography and its inherent limitations to make electrical contact to nanosystems. This idea is indicated schematically in Fig.~\ref{fig:INS}.

\subsection{Wireless interconnect to nano-sensors}

Another application is in the area of sensing. For example, nano-devices could be use as chemical and biological sensors, sensitive to their local chemical environment. A nanotube could be used as an antenna to couple to these nano-sensors, without the need for lithographically fabricated electronics.

\section{Two-nanotube transmission line properties}

In order to understand the nanotube antenna performance, we must first develop an RF circuit model for a transmission line consisting of two parallel nanotubes. We first review the RF circuit model for an individual nanotube, then discuss the equivalent circuit model for two spinless 1d wires, then discuss the circuit model appropriate for nanotubes, taking spin and band-structure degeneracy into account.

\subsection{Single nanotube RF properties}
In our recent work
\cite{2003:55-58,2002:129-144}, 
we considered the electrical properties of a SWNT above a ground plane in some detail. There, we found that, in addition to electrostatic capacitance and magnetic inductance, there were two additional distributed circuit elements to be considered: the quantum capacitance and the kinetic inductance. We briefly reiterate their physical origins here. 

The physical origin of the quantum capacitance comes from the finite density of states at the Fermi energy. In a quantum particle in a box, the spacing between allowed energy levels is finite. Because of this, to add an extra electron to the system takes a finite amount of energy above the Fermi energy. In one dimensional systems, this can be equated with an energy per unit length. From this energy per unit length, a capacitance per unit length can be calculated. From
\cite{2003:55-58,2002:129-144}, 
one finds the following expression for the (quantum) capacitance
per unit length:
\begin{equation}
{\mathcal C}_Q = {2 e^2\over h v_F}.
\end{equation}
The Fermi velocity for graphene and also carbon nanotubes is usually
taken as $v_F=8~10^5~m/s$, so that numerically
\begin{equation}
{\mathcal C}_Q = 100~aF/\mu m.
\end{equation}
The kinetic inductance has a simple physical origin as well. It is due to the charge-carrier inertia: electrons due not instantaneously respond to an applied electric field; there is some delay. For periodic electric fields, the electron velocity lags the electric field in phase, i.e. the current lags the voltage in phase. This appears as an inductance. It can be shown
\cite{2003:55-58,2002:129-144} 
that in 1d systems, this inductance is given by:\begin{equation}
\label{eq:lkinetic}
{\mathcal L}_K = {h\over 2 e^2 v_F},
\end{equation}
which comes out to be numerically
\begin{equation}
{\mathcal L}_K =  16~nH/\mu m.
\end{equation}
This simple model has been put on more rigorous grounds in\cite{2003:607-618}.

In a nanotube, there are four co-propagating quantum channels: two spin-up channels, and two spin-down channels. Each has its own kinetic inductance and quantum capacitance. All 4 channels have a common electrostatic capacitance to ground. This has a significant effect on the RF properties, as discuss in depth in \cite{2003:55-58,2002:129-144}.

\subsection{Circuit model for two 1d wires of spinless electrons}

\begin{figure}
\centering
\includegraphics[width=3.5in]{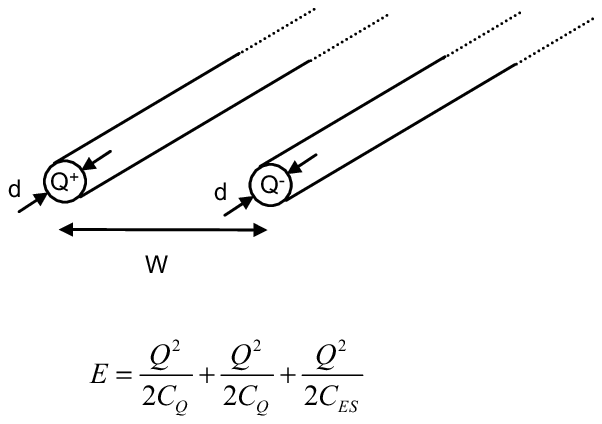}
\caption{Geometry of two-nanotube transmission line.}
\label{fig:twotubecartoon}
\end{figure}

In this section, we are interested in the differential modes of a two-nanotube transmission line system: There will be a voltage difference between the two nanotubes V(x,t), and a differential current I(x,t). For simplicity, consider two 1d wires of diameter d separated by a distance W, 
as shown in Fig.~\ref{fig:twotubecartoon}. Each wire has its own kinetic inductance per unit length; we neglect the magnetic inductance because the kinetic inductance dominates. 

For the two-nanotube transmission line, the issue of the quantum capacitance deserves some attention. 
We are interested in differential voltages and currents. If a voltage wave is excited, then there will be differentially large charge on one tube and smaller charge on the other tube. In the LL model, the energy cost to add charge is the same as to subtract charge. Thus, there is an energy cost for one tube to have excess charge, and there is an additional energy cost for the opposite tube to have decreased charge. In terms of the quantum capacitance, the energy cost for one tube to have excess charge Q is $\rho^2/C_Q$. Thus for a charge of $Q_+$ on one tube and a charge of $Q_-$ on the other tube, the net energy cost is $2 \rho^2/C_Q$. This can be equated to a capacitance between the tubes of $C_Q/2$ per unit length.

There will also be an electrostatic cross-capacitance between the wires, which is defined 
as the voltage drop from one tube to the other when there is an excess charge on one tube 
and a decreased charge on the other tube. This can be calculated from simple electrostatics to be:
\begin{equation}
\label{eq:celectrostatic}
{\mathcal C}_{ES} = { \pi \epsilon \over
  cosh^{-1}\bigl(W/d\bigr)}\approx {\pi\epsilon \over ln(W/d)},
\end{equation}

\begin{figure}
\centering
\includegraphics[width=3.5in]{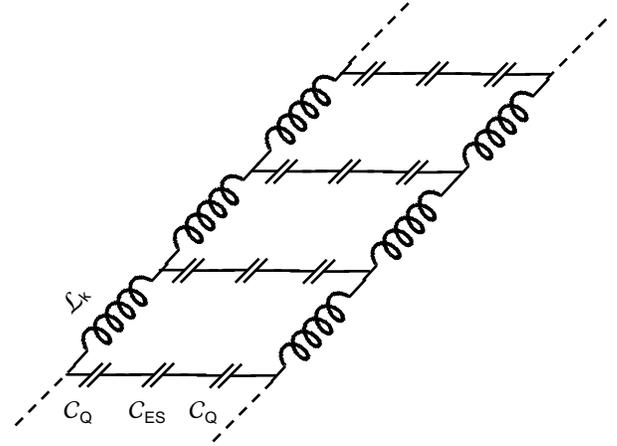}
\caption{RF circuit model for two 1d quantum wires with spinless electrons.}
\label{fig:twotubecircuit1}
\end{figure}

Since the energies add, the capacitances add inversely, i.e. the capacitances should be in series between the two nanotubes, as shown in Fig.~\ref{fig:twotubecartoon}. The effective circuit diagram is given in Fig.~\ref{fig:twotubecircuit1}. It is clear that the circuit model of Fig.~\ref{fig:twotubecircuit1} supports wave-like excitations of the current and voltage. Rather than calculating the wave properties explicitly, we move directly to the case of the two carbon nanotube transmission line.

\subsection{Circuit model for two carbon nanotubes}

\begin{figure}
\centering
\includegraphics[width=3.5in]{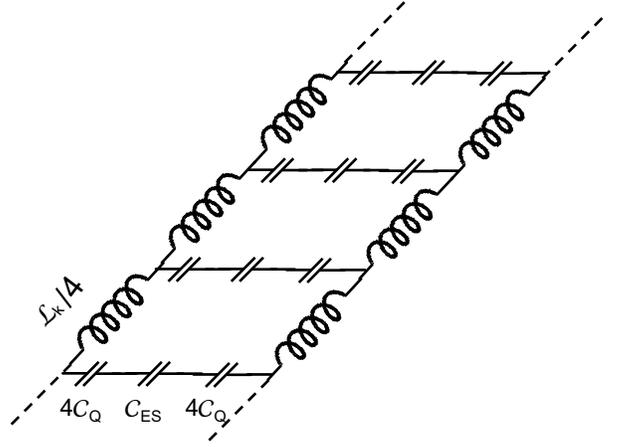}
\caption{RF circuit model for two-nanotube transmission line.}
\label{fig:twotubecircuit2}
\end{figure}

The circuit model for carbon nanotubes is more complicated, since each nanotube has four channels (two spin up, two spin down), each with its own kinetic inductance and quantum capacitance. For the differential mode excitations considered herein, the effective circuit model is modified. There are two spin orientations and two band structure channels that can propagate current, i.e. 4 1d quantum channels in parallel. Therefore, the kinetic inductance is 4 times lower than the one-channel case, and the quantum capacitance is 4 times higher than the one-channel case. The effective circuit diagram that takes these into account is given in Fig.~\ref{fig:twotubecircuit2}.

By simple applications of Kirchoff's laws to the circuit shown in Fig.~\ref{fig:twotubecircuit2}, we can come up with a differential equation for the differential voltage. If we write the voltage and current on tube 1 as $V_1$ and $I_1$, and the voltage and current on tube 2 as $V_2$ and $I_2$, and define the differential voltages and currents (assuming a harmonic time dependence) as:
\begin{eqnarray}
V_D&\equiv& V_1-V_2\\
I_D&\equiv& I_1-I_2
\end{eqnarray}
 then the following differential equation holds:
\begin{equation}
{\partial^2 V_{D}\over\partial x^2}-\gamma_p^2 V_D=0
\end{equation}
where the propagation constant is given by:
\begin{equation}
\gamma_p^2 \equiv 2({\mathcal R}+i\omega {\mathcal L}_K/4)(i\omega {\mathcal C}_{Total})
\end{equation}
Here we have introduced ${\mathcal R}$ as the resistance per unit length for a single tube in order to account for possible damping; this will be discussed in depth below. We use the subscript ``p'' for plasmon, because these excitations are collective oscillations of the 1d electron density, i.e. plasmons.

General solutions for the differential current and voltage can be written as:
\begin{eqnarray}
V_D(x)&=&V_0^+ e^{-\gamma_p x}+V_0^- e^{\gamma_p x}\\
I_D(x)&=&I_0^+ e^{-\gamma_p x}+I_0^- e^{\gamma_p x}\\
\label{eq:current}
&=&{V_0^+\over Z_C}e^{-\gamma_p x} - {V_0^-\over Z_C}e^{\gamma_p x},
\end{eqnarray}
where the characteristic impedance and wave velocity are given by:
\begin{eqnarray}
Z_{C}&\equiv&{1\over\sqrt{2}}\sqrt{{\mathcal R}+i\omega {\mathcal L}_K/4\over i\omega {\mathcal C}_{Total}}\\
v_p&\equiv&{1\over\sqrt{2}}{1\over{\sqrt{({\mathcal L}_K/4){\mathcal C}_{Total}}}},
\label{eq:vplasmon}
\end{eqnarray}
with
\begin{equation}
{\mathcal C}_{Total}^{-1}=(8{\mathcal C}_Q)^{-1} + {\mathcal C}_{ES}^{-1}.
\label{eq:ctotal}
\end{equation}
The characteristic impedance is so defined because the ratio of the voltage to the current is constant for a given propagation direction, i.e.
\begin{equation}
{V_0^+\over I_0^+}=-{V_0^-\over I_0^-}=Z_C.
\end{equation}
Numerically, for typical cases, we have:
\begin{eqnarray}
Z_{C}&\approx&{h\over 2 e^2} = 12~k\Omega\\
v&\approx&v_{Fermi}\approx 0.01~c.
\end{eqnarray}

Note that we are only considering differential mode here. There will also be common mode excitations, which will be wave-like as well. We do not discuss those here, nor do we claim that our circuit model is appropriate for common mode excitations.

\subsection{Discussion}

First, we neglected the magnetic inductance, which is justified because it is numerically much smaller than the kinetic inductance. 

Second, the wave velocity of this system is about 100 times smaller than the speed of light. This is because of the excess kinetic inductance. 

Third, the existence of wave-like current excitations is well-documented in the theoretical physics literature on 1d quantum systems\cite{1995:17040-17043,1996:10328-10331,1997:41-46,1998:1925-1928,1996:256-258,1996:L21-L26,1996:R5203-R5206,1998:1515-1526,2000:485-494}. Such slow-wave structures have been confirmed at GHz frequencies using 2d systems with kinetic inductance much larger than magnetic inductance by experiments performed by some of us\cite{2000:745-747,2004:03}. 

Fourth, our work is the first to consider the coupled nanotube transmission line. This should be put on more rigorous theoretical grounds by theoretical physicists, who would find an in phase and out of phase coupled charge oscillation mode. Similar work has been done already on closely spaced 2d electron gas systems\cite{1981:805-815}. Our circuit model describes the out of phase (differential) mode.

Fifth, the propagation constant and characteristic impedance are general expression that take into account loss. In the low loss case, they reduce to the more familiar forms. However, the expressions above are completely general including the case of high and low loss.

\section{Current distribution on a nanotube antenna}
\label{sec:currentdistsection}

\subsection{Qualitative discussion}

Fig.~\ref{fig:flaring} shows the standard, textbook geometry to calculate the current distribution on a classical, thin-wire antenna\cite{0471592684}. However, in our case, due to the kinetic inductance and quantum capacitance, the wave velocity is very different than a classical thin-wire antenna, where only the magnetic inductance is present.

In Fig.~\ref{fig:flaring}A, we consider the excited two-nanotube transmission line, where the ends are open. In this case, a standing-wave pattern is built up (as indicated) for the current and voltage along the two-nanotube transmission line. Because the currents on the two nanotubes are in equal and opposite direction, the far-field magnetic and electric fields generated by each of the wires individual cancels. Therefore, the radiated power is approximately zero.

In Fig.~\ref{fig:flaring}B, we consider ``flaring'' the two ends. If the flaring angle is small, the transmission line properties are almost the same, hence the standing wave pattern in the current is unchanged. However, because the wires are no longer close to each other at the ends, the far-field electric and magnetic fields generated by the wires near the end do not cancel, hence the system radiates power. Eventually, the flare angle becomes 90 degrees, and that is the geometry considered in this paper. The currents are quantitatively calculated below.

\subsection{2nd Order Flaring Effects}

The wave velocity for a traditional two-parallel wire transmission line is independent of the distance between the wires, and equal to the speed of light. Because of this, it is usually assumed that the current distribution for the flared two-nanotube system has the same wavelength as the unperturbed two-wire transmission line.

\begin{figure}
\centering
\includegraphics[width=3.5in]{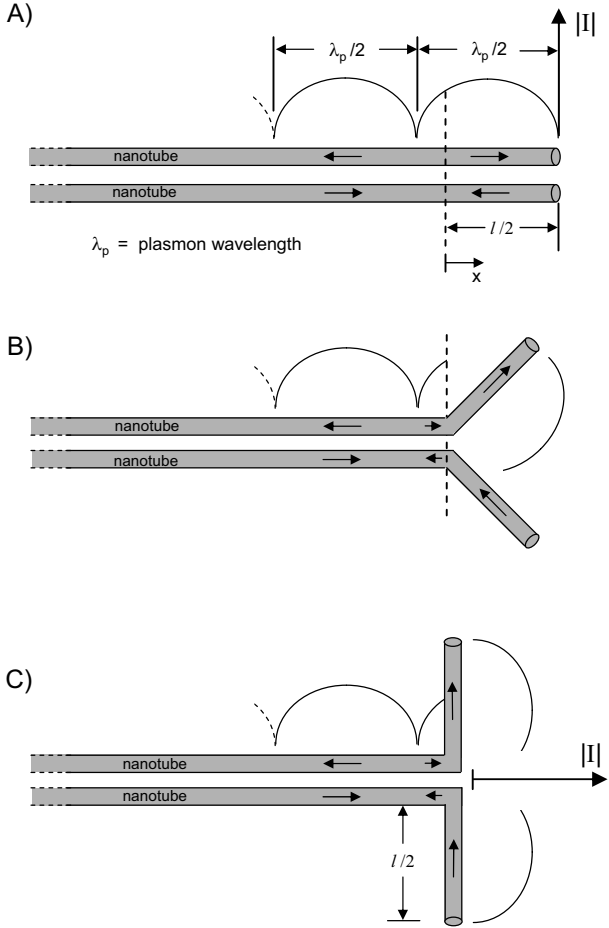}
\caption{Flaring. $\lambda_p$ is the plasmon wavelength, which is different than the free space wavelength $\lambda$.}
\label{fig:flaring}
\end{figure}

For two-wire nanotube transmission lines, the wave-velocity depends on the distance between the nanotubes. Therefore, the wave velocity for the flared nanotubes is different. The reason is simple: the kinetic inductance does not depend on the distance between the tubes, whereas the capacitance does. Therefore, the wave velocity $\sim 1/\sqrt{LC}$ depends on the distance between the tubes. This means, for the 90 degree flared nanotubes, that the wavelength of the current distribution is different. However, since the electrostatic capacitance is only sensitive to the log of the distance between nanotubes, this effect will be neglected herein. Therefore, we will assume that the current distribution of the flared nanotubes (Fig.~\ref{fig:flaring}C) is the same as that of the unflared nanotubes (Fig.~\ref{fig:flaring}A). We now calculate that current distribution.

\subsection{Quantitative prediction for arbitrary resistance}

We can use the transmission-line equations to develop expressions for the current distribution on the wire, explicitly and quantitatively including the effect of resistance along the tube, but neglecting (for the moment) the radiation resistance. If there is a positive-going voltage wave of amplitude $V_0^+$, it will be reflected off of the ends of the transmission line, with reflection coefficient 1 (since the ends are an open circuit). Therefore, there will be a negative going voltage wave of amplitude $V_0^-$ of equal amplitude. The propagation constant is $\gamma_p$. 

\begin{figure}
\centering
\includegraphics[width=3.5in]{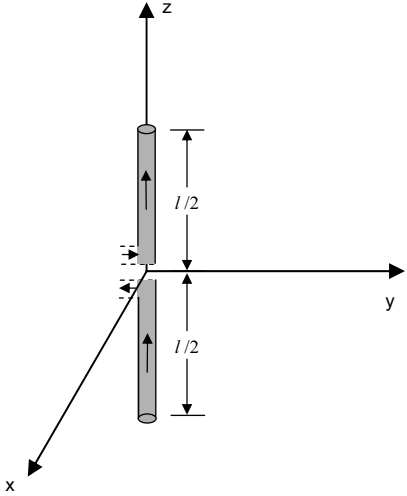}
\caption{Antenna geometry.}
\label{fig:antennageometry}
\end{figure}

The voltage along the antenna can be expressed then as
\begin{eqnarray}
V_D(x) &=&V_0^+e^{-\gamma_p (x-l/2)} + V_0^-e^{\gamma_p (x-l/2)}\\
&=& V_0^+e^{-\gamma_p (x-l/2)} + V_0^+e^{\gamma_p (x-l/2)}\\
&=& 2V_0^+cosh\bigl[\gamma_p (x-l/2)\bigr].
\end{eqnarray}
Note that, in this notation, the voltage at the terminals of the antenna $V_{terminals}$ is related to the amplitude of the positive going voltage wave $V_0^+$ by:
\begin{equation}
V_{terminals}=2 V_0^+ cosh\bigl[\gamma_p(l/2)\bigr]
\end{equation}
We can use equation~\ref{eq:current} to find the differential current on the line, and it is given by:
\begin{eqnarray}
I_D(x) &=&{V_0^+\over Z_C}e^{-\gamma_p (x-l/2)} -{V_0^-\over Z_C}e^{\gamma_p (x-l/2)}\\
&=&{V_0^+\over Z_C}e^{-\gamma_p (x-l/2)} -{V_0^+\over Z_C}e^{\gamma_p (x-l/2)}\\
&=&{2V_0^+\over Z_C}sinh\bigl[\gamma_p (l/2-x)\bigr].
\end{eqnarray}
This is referred to in Fig.~\ref{fig:flaring}, before flaring. After flaring, the detailed geometry is shown in Fig.~\ref{fig:antennageometry}. It is clear from inspection that both $I_1$ and $I_2$ are in the same direction (the positive z direction), and each will be equal to half of $I_D$. Therefore, the current on the active region of the antenna can be written as:
\begin{equation}
\label{eq:currentdistgeneral}
     \mbox{I}(z)=\left\{
     \begin{array}{rl}
{V_0^+\over Z_C} sinh\bigl[\gamma_p (l/2-z)\bigr] & 0<z<l/2 \\
{V_0^+\over Z_C}sinh\bigl[\gamma_p (z+l/2)\bigr] & 0>z>-l/2.
     \end{array} \right.
\end{equation}
These equations describe the nanotube antenna current distribution for arbitrary loss, neglecting the radiation resistance.

In the absence of loss (i.e. ${\mathcal R}=0$), the current can be written as:
\begin{equation}
\label{eq:currentdistnoloss}
     \mbox{I}(z)=\left\{
     \begin{array}{rl}
I_0 sin\bigl[k_p (l/2-z)\bigr] & 0<z<l/2 \\
I_0 sin\bigl[k_p (z+l/2)\bigr] & 0>z>-l/2.
     \end{array} \right.
\end{equation}
where $k_p$ is real, and equal to $\omega/v_p$, the plasmon velocity given by equation~\ref{eq:vplasmon}, 
and $I_0\equiv iV_0^+/Z_C$. This is what distinguishes a nanotube antenna from a traditional antenna, where the wavevector for the current is the same as the free-space wave-vector.

\begin{figure}
\centering
\includegraphics[width=3.5in]{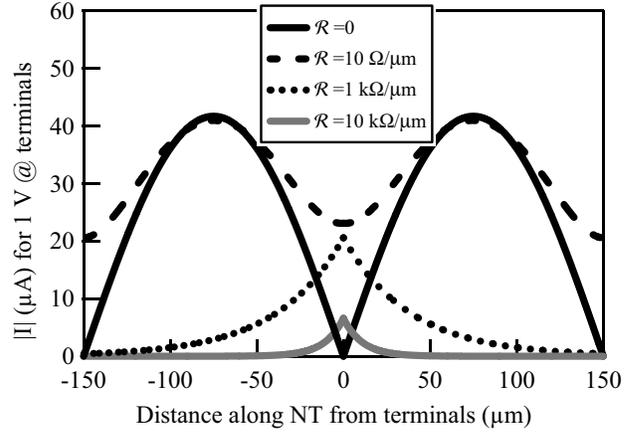}
\caption{Current Distribution for 1~V, 10 GHz excitation on a 300~$\mu$m antenna.}
\label{fig:currentdist}
\end{figure}

In Fig.~\ref{fig:currentdist}, we plot the magnitude of the ac current as a function of position for various values of R, for a 10 GHz frequency and 300~$\mu$m long nanotube antenna, for a voltage at the terminals of $V_{terminals}=1~V$. For low enough ${\mathcal R}$, the current distribution is approximately sinusoidal.  As ${\mathcal R}$ increases, the resonance behavior gives rise to an amplitude that decays exponentially 
with distance from the terminals.

\subsection{Effect of radiation on current distribution}

Below, we will calculate the far-field electric fields generated by the current distribution given in equations~\ref{eq:currentdistgeneral} and \ref{eq:currentdistnoloss}. This is not entirely self-consistent, because the current distribution expression given by equations~\ref{eq:currentdistgeneral} and \ref{eq:currentdistnoloss} neglect radiation. In reality, there will be a change in the current distribution due to the radiation. The far-field electric fields and the current distribution are related through a set of integro-differential equations, which can only be solved numerically. However, generally speaking for thin wire antennas, the current distribution is only slightly modified by the radiation, and is usually neglected. In this paper, we will assume the current distribution is not significantly changed by the radiation. This assumption should be put on more rigorous grounds in future work, but seems reasonable given what is known about traditional thin wire antennas.

In sum, we will neglect the effect of radiation on the current distribution and take equations~\ref{eq:currentdistgeneral} and \ref{eq:currentdistnoloss}
as a given for the rest of this paper, and turn to the consequences of this current distribution on radiation hence antenna properties.

\section{Radiation properties under no loss conditions}

Once the current distribution is known, we can treat each infinitesimal element of current as a radiator and add up (integrate) the contributions to determine the far-field electric field.
In this section, we will discuss the no-loss condition. While this may not be achievable in practice, it gives information about the ``ideal'' case. 

\subsection{Electric field}

Based on the known current distribution, it is straightforward to calculate the radiated electric and magnetic fields. We follow Balanis\cite{0471592684}. For a wire antenna of arbitrary current distribution $I_e$ along the z axis, the electric field in the far field region is in the $\theta$ direction, and given by:
\begin{equation}
\label{eq:radintegral}
E_\theta = i\eta {k e^{-ikr}\over 4\pi r}sin\theta \biggl[\int_{-l/2}^{+l/2}I(z)e^{ikzcos\theta}dz \biggr]
\end{equation}
Here $\eta$ is the characteristic impedance of free space, equal to $120\pi~\Omega$.

The key result of this paper is that, while for a traditional wire antenna 
the current distribution is periodic with wave-vector given by 
$k=2\pi/ \lambda$, where $\lambda$ is the free space electromagnetic wavelength, 
for a nanotube the current is periodic with wavevector given by 
$k=2\pi/ \lambda_p$, where $\lambda_p$ is the plasmon wavelength.
This causes the integral of equation~\ref{eq:radintegral} to be different from a traditional thin-wire antenna.

Numerically, $k_p\approx 100 k$, i.e. $\lambda_p\approx {1\over 100}\lambda$, where $\lambda$ is the free space wavelength. For simplicity, we will assume this relationship to be exact from now on.

 We can write the electric field then as:
\begin{eqnarray}
\label{eq:radintegrallowloss}
\lefteqn{E_\theta= i\eta {k e^{-ikr}\over 4\pi r}sin\theta } \nonumber \\
& &X \,\Biggl\lbrace\int_{-l/2}^{0}   I_0 sin\biggl[k_p\Bigl({l\over2}+z\Bigr)\biggr] e^{ikz cos\theta}dz \nonumber \\
& &  \,\,\,\,\,\,\,+ \int_{0}^{+l/2} I_0 sin\biggl[k_p\Bigl({l\over2}-z\Bigr)\biggr]   e^{ikz cos\theta}dz \Biggr\rbrace
\end{eqnarray}

Equation~\ref{eq:radintegrallowloss} can be evaluated, and the result is:
%\begin{eqnarray}
%\lefteqn{E_\theta=i\eta{k\over k_p} {I_0 e^{-ikr}\over 2\pi r}X}\nonumber\\ 
%&&sin \,\theta {cos\biggl({kl\over2}cos\theta\biggr) - cos\biggl({k_p l\over 2}\biggr)\over 1 - (k/k_p)^2cos^2\theta}
%\end{eqnarray}

\begin{equation}
\label{eq:eplaneradpat}
E_\theta=i\eta{k\over k_p} {I_0 e^{-ikr}\over 2\pi r} sin \,\theta {cos\Bigl({kl\over2}cos\theta\Bigr) - cos\Bigl({k_p l\over 2}\Bigr)\over 1 - (k/k_p)^2cos^2\theta}
\end{equation}

%Since numerically $k/k_p \approx 0.01$, we can ignore the denominator in the brackets to a good approximation (0.1\%), and write the power as:

%\begin{equation}
%E_\theta\approx i\eta{k\over k_p} {I_0 e^{-ikr}\over 2\pi r} sin \,\theta {cos\Bigl({kl\over2}cos\theta\Bigr) - cos\Bigl({k_p l\over 2}\Bigr)}
%\end{equation}

%\begin{eqnarray}
%\lefteqn{E_\theta=i\eta{k\over k_p} {I_0 e^{-ikr}\over 2\pi r}X}\nonumber\\ 
%&&{sin \,\theta\over 1 - (k/k_p)^2cos^2\theta}\biggl[cos\biggl({kl\over2}cos\theta\biggr) - cos\biggl({k_p l\over 2}\biggr)\biggr]
%\end{eqnarray}

\subsection{Poynting vector and radiation intensity}

The Poynting vector can be calculated from the electric field, resulting in:

%\begin{eqnarray}
%\lefteqn{W_{av}=\hat r {1\over 2 \eta} |E_\theta|^2 =\biggl({k\over k_p}\biggr)^2{I_0^2 \eta\over 8 \pi^2 r^2}X}\nonumber  \\ 
%&& {sin^2\theta\over[ 1-(k/k_p)^2 cos^2\theta]^2}\biggl[cos\biggl({kl\over 2}cos\theta\biggr) - cos\biggl({k_p l\over 2}\biggr)\biggr]^2
%\end{eqnarray}

\begin{eqnarray}
\label{eq:poynting}
\lefteqn{\vec W_{av}=\hat r {1\over 2 \eta} |E_\theta|^2 =}\\
&&\biggl({k\over k_p}\biggr)^2{I_0^2 \eta\over 8 \pi^2 r^2}\Biggl[sin \,\theta {cos\Bigl({kl\over2}cos\theta\Bigr) - cos\Bigl({k_p l\over 2}\Bigr)\over 1 - (k/k_p)^2cos^2\theta}\Biggr]^2
\end{eqnarray}

\begin{figure}
\centering
\includegraphics[width=3.5in]{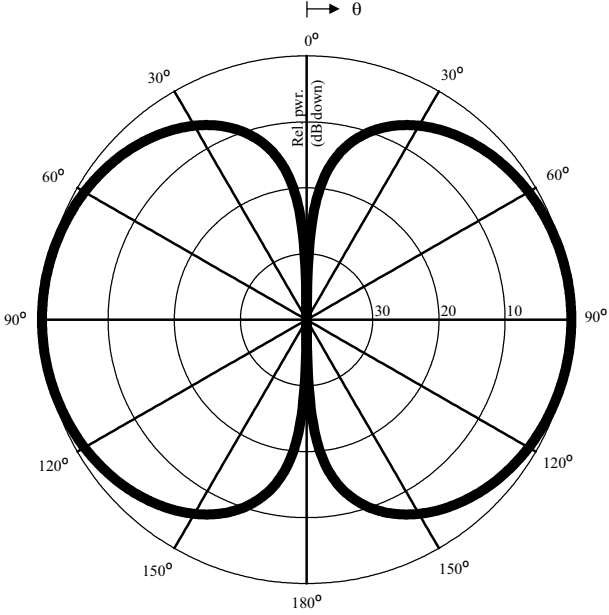}
\caption{E-plane antenna pattern, based on equation~\ref{eq:eplaneradpat}. We assume $k_p l/2=\pi$.}
\label{fig:antpat}
\end{figure}

Next, the (time-average) radiation intensity can be written as

\begin{eqnarray}
\label{eq:radintensity}
\lefteqn{U=r^2 W_{av} =}\\
&&\biggl({k\over k_p}\biggr)^2{I_0^2 \eta\over 8 \pi^2}\Biggl[sin \,\theta {cos\Bigl({kl\over2}cos\theta\Bigr) - cos\Bigl({k_p l\over 2}\Bigr)\over 1 - (k/k_p)^2cos^2\theta}\Biggr]^2
\end{eqnarray}
The antenna pattern is similar to a regular wire antenna as long 
as the length is not near $l/\lambda = 0.02~n$, with n an integer. If the length is near one of these multiples, the pattern develops extra lobes. The radiation pattern is plotted in Fig.~\ref{fig:antpat}, assuming $l/\lambda = 0.01$, i.e. $l/\lambda_p = 1$.

\subsection{Total radiated power}

The total radiated power can be determined by integrating the radiation intensity over a sphere, i.e.

\begin{eqnarray}
\lefteqn{P_{rad}=\oint\limits_S \vec W_{av}\cdot d\vec s=\int\limits_0^{2\pi}\!\int\limits_0^\pi W_{av}r^2 \,sin\theta \,d\theta \,d\phi}\nonumber\\
&=&\biggl({k\over k_p}\biggr)^2{I_0^2 \eta\over 4 \pi}\int\limits_0^\pi sin^3\,\theta\Biggl[{cos\Bigl({kl\over2}cos\theta\Bigr) - cos\Bigl({k_p l\over 2}\Bigr)\over 1 - (k/k_p)^2cos^2\theta}\Biggr]^2\,d\theta\\
&=&\biggl({k\over k_p}\biggr)^2{I_0^2 \eta\over 2}\xi(k l,k_p l),
\end{eqnarray}

where we have defined the function $\xi(k l,k_p l)$ as
\begin{equation}
\xi(k l,k_p l)\equiv{1\over 2 \pi}\int\limits_0^\pi sin^3\,\theta\Biggl[{cos\Bigl({kl\over2}cos\theta\Bigr) - cos\Bigl({k_p l\over 2}\Bigr)\over 1 - (k/k_p)^2cos^2\theta}\Biggr]^2\,d\theta.
\end{equation}
The function $\xi(k l,k_p l)$ can be evaluated numerically. The function $\xi$ is of order unity, and plotted in the appendix. It is periodic with $k_p l$.

\subsection{Radiation resistance}
\label{sec:radressec}

The radiation resistance is defined by
\begin{equation}
\label{eq:rradlowloss}
R_r \equiv {2 P_{rad}\over |I_0|^2 }=\biggl({k\over k_p}\biggr)^2{\eta}~\xi(k l,k_p l).
\end{equation}
Since $\xi$ is of order unity, the radiation resistance is of order $\eta/10^4=0.04~\Omega$.

\subsection{Directivity and effective aperture}

The directivity is given by the maximum value of the radiated intensity divided by its average value, i.e.:
\begin{equation}
D={W_{av}|_{max}/A \over  \oint\limits_S \vec W_{av}\cdot d\vec s}
\end{equation}
We have numerically evaluated D as a function of $kl$ and find that it is 2 as long as the length is not near $l/\lambda = 0.02~n$, with n an integer. The pattern is that of a simple thin-wire dipole radiator.
If the length is near one of these multiples, the pattern develops extra lobes, and in that case the 
directivity can be as high as 5-6. In the appendix, we plot the directivity determined numerically as a function of $kl$.

The effective area is related to the directivity through:
\begin{equation}
A_{eff}={\lambda ^2\over 4 \pi}D
\end{equation}
Therefore, for most values of $kl$, the effective area is approximately given by:
\begin{equation}
A_{eff}\approx{\lambda ^2\over 2 \pi}
\end{equation}
This is similar to a thin-wire antenna.

\subsection{Input impedance}

The radiation resistance relates the power radiated to $I_0$, the maximum amplitude of the current along the nanotube. However, the current at the terminals is equal to $I_0~sin(k_pl/2)$. The input resistance due to radiation $R_{in}$ is related to the power dissipated due to radiation through $P_{rad}=I^2 R_{in}$, where $R_{in}$ is the current at the terminals.
Therefore, taking this into account, the input resistance (when there is no intrinsic loss) is given by:
\begin{equation}
R_{in}={ R_r \over sin^2\bigl(k_p l/ 2)  }
\end{equation}
The input reactance is easy to understand from Fig.~\ref{fig:flaring} as the input impedance of a two-nanotube transmission line of length l/2. This is given by:
\begin{equation}
\label{eq:inputreactance}
Z_{in} = - i Z_c cot (k_p l/2).
\end{equation}
There will also be an input reactance due to energy radiated and absorbed. However, numerically this will not be as large as the input reactance of equation~\ref{eq:inputreactance}, so is neglected.

\begin{figure}
\centering
\includegraphics[width=3.5in]{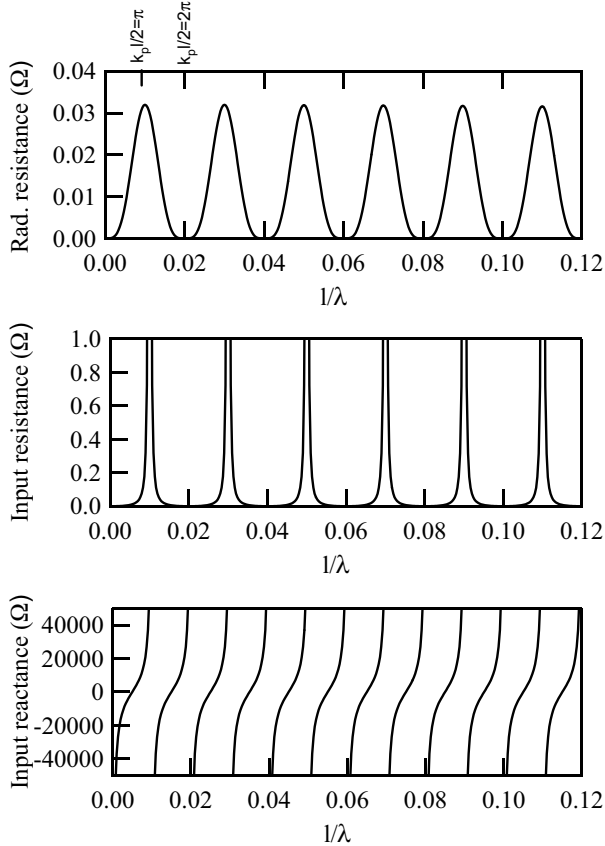}
\caption{Plot of radiation resistance, input resistance, and input reactance as a function of $l/\lambda$, where $\lambda$ is the free space wavelength, assuming that $k_p=100 k$. The x-axis is $l/\lambda = kl/2\pi = 0.01 k_pl/2 \pi$.}
\label{fig:radresistance}
\end{figure}

In Fig.~\ref{fig:radresistance}, we plot the radiation resistance, input resistance, and input reactance as a function of $l/\lambda$, where $\lambda$ is the free space wavelength, assuming that $k_p=100 k$.

\subsection{Discussion}

\begin{figure}
\centering
\includegraphics[width=3.5in]{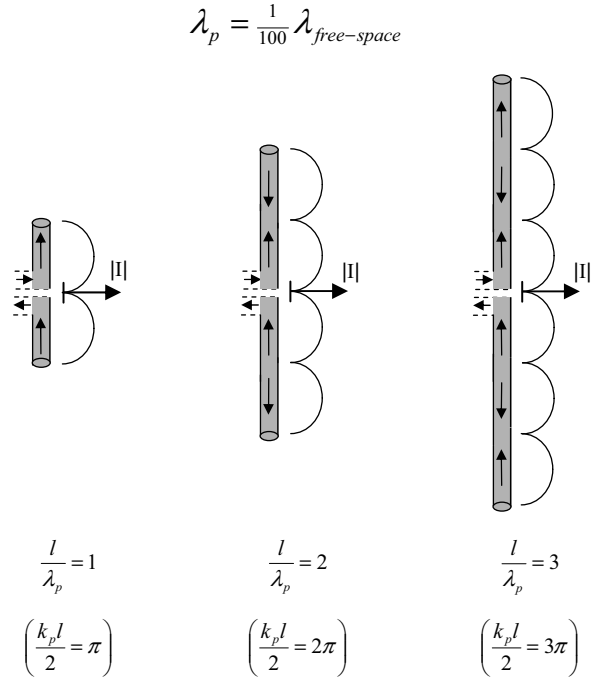}
\caption{Current distribution schematic for various lengths.}
\label{fig:antArray}
\end{figure}

Why is the radiation resistance so low, and why is it periodic in $k_p l/2$?  The answer to that question is quite simple. To illustrate, we show schematically in Fig.~\ref{fig:antArray} the current pattern on the antenna for three different values of $k_p l/2$.

The elements form what can be considered a phased array of current sources, but each element is out of phase with its nearest neighbor by 180 degrees. The far-field electric field is the sum of the fields generated by each element. Because each element is out of phase, the fields from the individual elements cancel if there is an even number of elements. If there is an odd number of elements, all but one of the elements cancel.

This analysis can be carried even further. The analysis suggests that one can neglect all but the last odd element as radiating. This suggests that the antenna properties of a nanotube antenna whose length is an odd integer multiple of the $\lambda_p$ is equivalent to short thin wire antenna of length $\lambda_p$. 
Indeed, this is true quantitatively. The radiation resistance of a thin-wire short antenna is given by:
\begin{equation}
\label{eq:rradthinwire}
R_{rad}=80 \pi^2 \biggl({l\over \lambda} \biggr)^2
\end{equation}
If we take $l=\lambda_p$, we get a radiation resistance of $0.08~\Omega$, which is almost exactly that predicted by our theory in Fig.~\ref{fig:radresistance}. There is a factor of 2 difference because, for the last odd element, the current is not constant but periodic in space, whereas equation~\ref{eq:rradthinwire} for a thin-wire antenna assumes a constant spatial current distribution. The average current for a sinusoidal current distribution is  $1/\sqrt{2}$ smaller than the max, thus accounting for the factor of 2 difference.

\begin{figure}
\centering
\includegraphics[width=3.5in]{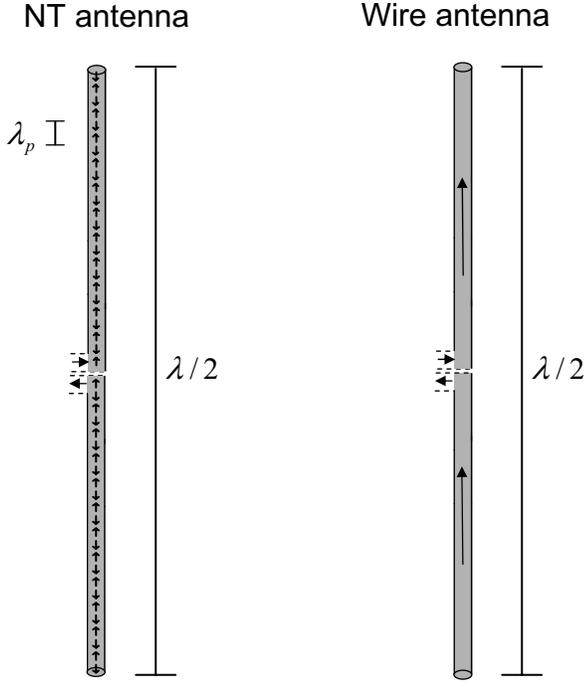}
\caption{Current distribution for nanotube vs. wire antenna for length $\lambda/2$.}
\label{fig:array2}
\end{figure}

For longer antennas, where the length is comparable to the free-space wavelength, the elements of length $\lambda_p$ are no longer close to each other compared to the free space wavelength, so that the situation is quantitatively more complicated. Equation~\ref{eq:rradlowloss} quantifies this situation. To emphasize this situation, in Fig.~\ref{fig:array2} we show a schematic comparison of the current distribution on a thin-wire antenna of length $\lambda/2$ vs. the current distribution on a nanotube antenna of the same length.

This analysis suggests then, for a lossless nanotube, that making an antenna with length longer that one plasmon wavelength is not at all beneficial in terms of antenna properties. Since the plasmon wavelength is short, a nanotube antenna will in the best case scenario be only as good as a short thin-wire antenna, with length given by the plasmon wavelength, which is about 100 times smaller than the free space wavelength. This, presumably, is a general property of slow-wave antennas, of which a nanotube is an extreme example.

\section{Radiation properties under arbitrary loss conditions}

\subsection{Electric field}

In the above section~\ref{sec:currentdistsection}, we solved for the current distribution under arbitrary loss conditions. 
We now use this current distribution to calculate the radiation properties. On substituting the general expression for the current (equation~\ref{eq:currentdistgeneral}) into equation~\ref{eq:radintegral} for determining the far-field electric field, we have:
\begin{eqnarray}
\lefteqn{E_\theta= i\eta {k e^{-ikr}\over 4\pi r}sin\theta } \nonumber \\
& &X \,\Biggl\lbrace\int_{-l/2}^{0}   {V_0^+ \over Z_C} sinh\biggl[\gamma_p\Bigl({l\over2}+z\Bigr)\biggr] e^{ikz cos\theta}dz \nonumber \\
& &  \,\,\,\,\,\,\,+ \int_{0}^{+l/2} {V_0^+ \over Z_C} sinh\biggl[\gamma_p\Bigl({l\over2}-z\Bigr)\biggr]   e^{ikz cos\theta}dz \Biggr\rbrace
\label{eq:efieldgeneral}
\end{eqnarray}
Equation~\ref{eq:efieldgeneral} can be evaluated, and the result is:
\begin{equation}
E_\theta=-i\eta{k\over \gamma_p} {V_0^+\over Z_C} {e^{-ikr}\over 2\pi r} sin \,\theta {cos\Bigl({kl\over2}cos\theta\Bigr) - cosh\Bigl({\gamma_p l\over 2}\Bigr)\over 1 + (k/\gamma_p)^2cos^2\theta}
\end{equation}

\subsection{Poynting vector and radiation intensity}

The Poynting vector can be calculated from the electric field. Because $\gamma_P$ and $Z_C$ are complex, the result cannot easily be simplified, but can be numerically calculated from equation~\ref{eq:poynting}. The radiation intensity can be numerically calculated from equation~\ref{eq:radintensity}. The total radiated power can be determined by numerically integrating the radiation intensity over a sphere. The radiation resistance is not meaningful when there is intrinsic loss distributed along the antenna. The radiation pattern and directivity can also be numerically calculated.

\subsection{Input impedance}

If we neglect the energy radiated, the input impedance is given by:
\begin{equation}
\label{eq:zinput}
Z_{in} = Z_c coth (\gamma_p l/2).
\end{equation}
This can also be numerically calculated.

\section{Classification of loss regimes}

Loss, or resistance, is an important parameter in nanotube antennas. There are two ways in which low-loss can be defined. The first is that the frequency is high enough, and the loss low enough, so that the wave propagation on the two-nanotube transmission line is dispersion-free. Mathematically, the requirement for this is:
\begin{equation}
\label{eq:lowlossdef1}
i\omega {\mathcal L}_K/4 >> {\mathcal R}
\end{equation}
If one uses our recently measured value of $10~k\Omega/\mu m$, the low loss condition translates into a frequency requirement of:
\begin{equation}
f>400~GHz
\end{equation}
However, if lower resistance tubes are grown this could be lower. For example, at cryogenic temperatures this requirement may be different. This issue is further discussed below.
Under these low-loss conditions, the wave-vector is given by:
\begin{equation}
\gamma_p \approx ik_p + \alpha.
\end{equation}
The attenuation constant $\alpha$ is given by:
\begin{equation}
\label{eq:alphaequals}
\alpha = {1\over 2} {{\mathcal R}\over Z_C}
\end{equation}
The physical interpretation of the attenuation coefficient is the length over which a propagating wave on the two-nanotube transmission line decays in amplitude by $1/e$.
Additionally, under these conditions (equation~\ref{eq:lowlossdef1}), the characteristic impedance is real and given by:
\begin{equation}
Z_c\approx \sqrt{{\mathcal L}_K\over 8 {\mathcal C}_{Total}},
\end{equation}

A second, stricter definition of low-loss requires equation~\ref{eq:lowlossdef1} be satisfied and, in addition, 
requires that the wave is not significantly attenuated over the length of the antenna. In mathematical terms, this can be expressed as the condition:
\begin{equation}
\label{eq:lowlossdef2}
{\alpha l\over 2} << 1
\end{equation}
This second condition depends on the length of the nanotube.

\section{Low loss calculations}

In this section, we are interested in determining the effect of loss on antenna performance in the low-loss regime defined as both eqns.~\ref{eq:lowlossdef1} and~\ref{eq:lowlossdef2}. We seek to find expressions for antenna efficiency correct to linear order in $\alpha L/2$.

For the following, we will divide the discussion into two cases: on-resonance, defined by $k_p l/2=n\pi$, and off resonance, defined by $k_p l/2 \not= n\pi$.

\subsection{Resonance condition}

Since the radiation resistance (which is already quite low) is highest on resonance, this case would be the most logical case for maximizing the antenna efficiency.

\subsubsection{Resistive losses}

Neglecting radiation losses, what is the loss due to the ohmic dissipation? This depends on the resistance per length, antenna length, and frequency. We first calculate the input impedance (neglecting radiation resistance) on resonance. One can show that, on resonance, the real part of the input impedance (equation~\ref{eq:zinput}) is given approximately by:
\begin{equation}
\label{eq:rezinlowloss}
Re(Z_{in}) \approx {4 Z_c^2\over {\mathcal R} l}
\end{equation}
This allows us to calculate the ohmic losses in the ultra-low loss condition. Specifically, for a given voltage at the terminals $V_{terminals}$, the power dissipated due to ohmic losses is given by:
\begin{equation}
\label{eq:pohmiclowloss}
P_{ohmic}={1\over 2}{|V_{terminals}|^2\over Re(Z_{in})}={1\over 2}|V_{terminals}|^2{{\mathcal R}l\over 4 Z_c^2}
\end{equation}
This will be used later.

Note that the imaginary part of the input impedance on resonance will be zero in the presence of any small amount of loss.

\subsubsection{Radiative losses}

We seek an expression for the radiated power as a function of the voltage applied at the terminals, in order to compare with equation~\ref{eq:pohmiclowloss}.

In the ultra-low loss case, the overall current distribution (equation~\ref{eq:currentdistnoloss}) is not significantly altered, and therefore the radiation resistance is not significantly altered. For the purposes of the radiation resistance it is sufficient to assume that the current distributing is still approximately sinusoidal with amplitude $I_0$.
If we make this assumption, then the power dissipated due to radiation is still related to the radiation resistance through equation~\ref{eq:rradlowloss}, i.e.:
\begin{equation}
\label{eq:praddef}
P_{rad} = {1\over 2}{|I_0|^2  R_r}
\end{equation}
On resonance, the radiation resistance $R_r$ is approximately equal to $0.03~\Omega$, as discussed above in section~\ref{sec:radressec}.

When there are resistive losses, the current distribution is slightly modified from being sinusoidal and the current at the terminals on resonance is given by $I_0 sinh(\gamma_p l/2)$. On resonance, the current at the terminals can be approximated as 
\begin{equation}
I_{terminals}\approx 2 I_0 (\alpha l/2)
\end{equation}
(The factor of 2 comes because the differential current at the terminals is 2 times the current on an individual tube.) 
Therefore, the radiated power can be written as:
\begin{equation}
P_{rad} = {1\over 2}{|I_0|^2  R_r}={1\over 2}{|I_{terminals}|^2\over  (\alpha l)^2 } R_r
\end{equation}
%Furthermore, the input resistance is related to the radiation resistance by requiring the power dissipated due to radiation (equation ?) be equal to the $2 I_{terminals}^2/R_in$, i.e.
%\begin{equation}
%{1\over 2}{|I_0|^2 R_r}= {1\over 2}{ |I_{terminals}|^2 R_{in}}
%\end{equation}
%This implies, then, that the input resistance is given by:
%\begin{eqnarray}
%R_{in}&=&{R_r\over (\alpha l/2)^2}\\
%&=&{R_r\over 4} \biggl({4 Z_c\over {\mathcal R} l}\biggr)^2
%\end{eqnarray}
This allows us to calculate the power dissipated as radiation. Specifically, for a given voltage at the terminals $V_{terminals}$, the power dissipated due to radiation is given by:
\begin{equation}
\label{eq:pradlowloss}
P_{rad} ={1\over 2}{|I_{terminals}|^2\over (\alpha l)^2 } R_r
={1\over 2}\biggl|{V_{terminals}\over Re(Z_{in})}\biggr|^2 {1\over (\alpha l)^2}R_r
\end{equation}
Based on equation~\ref{eq:rezinlowloss} the above equation~\ref{eq:pradlowloss} can be expressed as:
\begin{equation}
\label{eq:pradlowloss2}
P_{rad}={1\over 2}|V_{terminals}|^2 {R_r\over Z_c^2}{1\over 4}
\end{equation}
This will be used later.

A comment about expression~\ref{eq:pradlowloss2}: This is interesting, because it shows that the radiated power does not depend on the resistance per length of the nanotube, in the low-loss limit. This is because the current distribution is assumed to be the same regardless of the loss, which is approximately true. The radiated power depends only on the current distribution, so does not change in this approximation. Only the current at the terminals changes, and only by a small amount in the low-loss approximation.

\subsubsection{Antenna efficiency}
\begin{figure}
\centering
\includegraphics[width=3.5in]{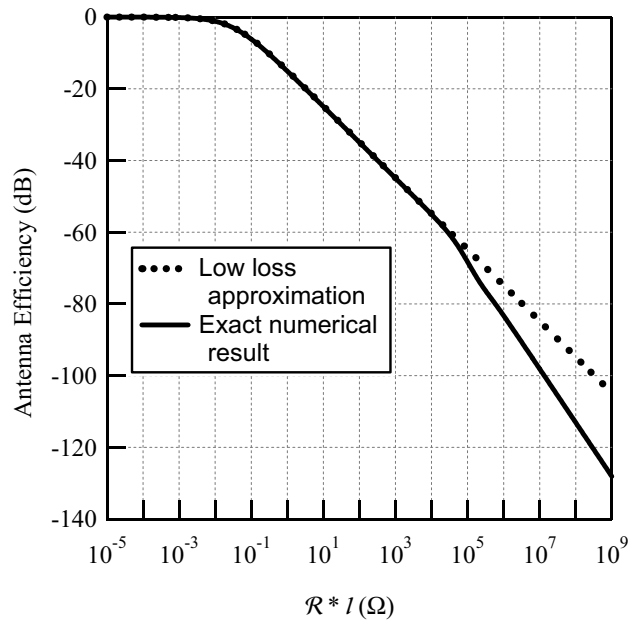}
\caption{Antenna efficiency vs. ${\mathcal R} l$, assuming $l=0.01\lambda$. The result is independent frequency or length, as long as the equation is true.}
\label{fig:anteff}
\end{figure}

We define the antenna efficiency as the ratio of the power dissipated in radiation to the total power dissipated (radiation and ohmic), i.e.
\begin{equation}
A.E.\equiv {P_{rad}\over P_{rad}+P_{ohmic}}
\end{equation}
Based on the above equations~\ref{eq:pradlowloss2} and~\ref{eq:pohmiclowloss}, this can be written as:
\begin{equation}
\label{eq:aelowloss}
A.E.={1\over 1+{\mathcal R}l/ R_r}={1\over 1+{\mathcal R}l/ 0.03~\Omega}
\end{equation}
This means that as soon as the nanotube dc resistance of length l exceeds $0.03~\Omega$, the resistive losses dominate. We plot in Fig.~\ref{fig:anteff} the antenna efficiency as a function of $R l$. Numerically, expression~\ref{eq:aelowloss} is good to within 0.3 dB if ${\mathcal R}l$ is less than $10^4~\Omega$.

\begin{figure}
\centering
\includegraphics[width=3.5in]{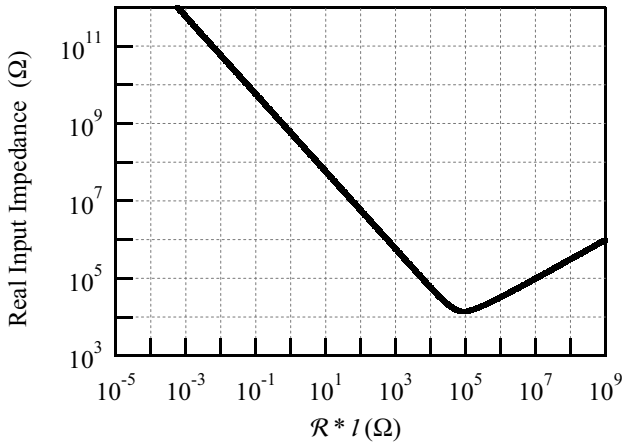}
\caption{Real input impedance at antenna terminals vs. ${\mathcal R} l$, assuming $l=0.01\lambda$. The result is independent frequency or length, as long as the equation is true.}
\label{fig:rinvsreff}
\end{figure}

\subsection{Off resonance condition}

In the off resonance case, the discussion is similar. We want to determine the power dissipated due to 
radiation losses and ohmic losses, then calculate the antenna efficiency.

\subsubsection{Resistive losses}

For small damping, off resonance, the real part of the the input impedance
 (equation~\ref{eq:zinput}) is given approximately by:
\begin{equation}
\label{eq:rezinoffres}
Re(Z_{in}) = Z_c {\alpha l/2\over sin^2(k_p l/2)}={{\mathcal R} l/4 \over  sin^2(k_p l/2)}
\end{equation}
This allows us to calculate the ohmic losses in the ultra-low loss condition. For a given voltage at the terminals $V_{terminals}$, the power dissipated due to ohmic losses is given by:
\begin{equation}
\label{eq:pohmicoffres}
P_{ohmic}={1\over 2}{|V_{terminals}|^2\over Re(Z_{in})}={1\over 2}|V_{terminals}|^2{{sin^2(k_p l/2)\over {\mathcal R} l/4}}
\end{equation}

\subsubsection{Radiative losses}

Off resonance, the current at the terminals is not near a null, so that the sinusoidal approximation can be used, i.e.:
\begin{equation}
I_{terminals}=2~I_0 sin(k_pl/2)
\end{equation}
Combining this with equation~\ref{eq:praddef}, we find:
\begin{eqnarray}
\label{eq:pradoffres}
  P_{rad} &=&{1\over 2}{|I_{terminals}|^2\over 4~ sin^2(k_p l/2) } R_r\\
&=&{1\over 2}\biggl|{V_{terminals}\over Re(Z_{in})}\biggr|^2 
{1\over 4~ sin^2(k_p l/2) } R_r\\
%&=&{1\over 2}|V_{terminals}|^2{1\over 4~ sin^2(k_p l/2) }{1\over |Re(Z_{in})|^2} R_r\\
%&=&{1\over 2}|V_{terminals}|^2{1\over 4~ sin^2(k_p l/2) } \biggl({{sin^2(k_p l/2)\over {\mathcal R} l/4}}\biggr)^2 R_r\\
&=&{1\over 2}|V_{terminals}|^2 sin^2(k_p l/2){4 R_r \over ({\mathcal R}l)^2  },
\end{eqnarray}
where we have used equation~\ref{eq:rezinoffres}. This will be used later.

\subsubsection{Antenna efficiency}

Based on the above equations~\ref{eq:pradoffres} and~\ref{eq:pohmicoffres}, this can be written as:
\begin{equation}
A.E.={1\over 1+{\mathcal R}l/ R_r}
\end{equation}
This is the same as the on-resonance condition. However, the radiation resistance $R_r$ is frequency dependent, and maximum on resonance. (See Fig.~\ref{fig:radresistance}.) Therefore, the antenna efficiency will be maximum on resonance.

\section{High loss calculations}

\subsection{Loss classification}

In the above section, we calculated the antenna efficiency in the low-loss case according to the criteria $\alpha l/2 < 1$. In this section, we seek to determine the antenna efficiency in the high loss case according to the criteria $\alpha l/2 > 1$. If this criteria is met, and if the antenna is designed for microwave frequencies, then it is also going to be true that the system is in the high-loss case according to the condition
$\omega {\mathcal L}_K < {\mathcal R}$. We elaborate.

If the antenna is designed for microwave frequencies, then the length will be of order the plasmon wavelength at microwave frequencies, which is of order 100~$\mu$m. If this is the case, and it is true that $\alpha l/2 > 1$, then according to equation~\ref{eq:alphaequals}, the resistance per length will be at least of order 100~k$\Omega/\mu$m, which is numerically larger than
$\omega {\mathcal L}_K$ at microwave frequencies. Therefore, the high-loss calculations to be discussed in the section will be high loss in both senses (eqns.~\ref{eq:lowlossdef1} and~\ref{eq:lowlossdef2}).

\subsection{Qualitative discussion}

In the high loss case, the current distribution is dramatically changed. Essentially, the only spatial region of the antenna which carries current is the region within $\alpha^{-1}$ of the terminals. By definition $\alpha l/2 > 1$, this is a small fraction of the entire antenna. This is seen clearly in Fig.~\ref{fig:currentdist}, where we plot the current distribution for various loss values. For the highest values of ${\mathcal R}$, the current flows only near the terminals. It is dissipated as ohmic losses before reaching the end of the nanotubes, which are far away from the terminals on the scale of $\alpha^{-1}$. Therefore, the radiated power and radiation efficiency will be significantly {\it lower} than the low-loss prediction, equation~\ref{eq:aelowloss}.

\subsection{Numerical calculation}

We have numerically evaluated the radiated power and ohmic dissipated power as a function of the ${\mathcal R} l$ product, and then calculated the antenna efficiency as defined above. The radiated power is calculated numerically by integrating the radiation intensity calculated from the electric field (equation~\ref{eq:efieldgeneral}) over a sphere. The numerical integration was performed using a simple script in Mathematica. A length $l=0.01~\lambda$ was assumed for the calculations, which corresponds to the on-resonance case in the low-loss condition as discussed above.

The ohmic losses are calculated numerically by calculating the power according to:
\begin{equation}
P_{ohmic}={1\over 2} Re\bigl[V_{terminals}I^*_{terminals}\bigr],
\end{equation}
and exploiting the input impedance given in equation~\ref{eq:zinput} for the relationship between $V_{terminals}$ and $I_{terminals}$.

In Fig.~\ref{fig:anteff}, we plot the exact numerical solution for the antenna efficiency as a function of the parameter ${\mathcal R} l$. Interestingly, this curve is universal regardless of the frequency, the numerical value of $\mathcal{R}$, or the numerical value of $l$, as long as one assumes $l=0.01\lambda$. Since we showed in the low-loss case that the antenna efficiency is maximized on resonance (which is true of $l=0.01\lambda$), it is reasonable to assume antenna operation in the high loss case would be for the same length.

\subsection{Discussion}

From Fig.~\ref{fig:anteff}, it can be seen that the low-loss approximation breaks down for an ${\mathcal R}l$ value of around 20~k$\Omega$. For a nanotube antenna with a microwave resonance frequency, the length will be of order 100~$\mu$m. For this antenna, the low-loss approximation will break down at a numerical value of ${\mathcal R}\sim 100~\Omega$. This value is about 100 times lower than our measured value of ${\mathcal R}$ of 10~k$\Omega/\mu m$. Therefore, for available nanotube technology, the antenna will most likely operate in the high-loss region. However, since long-nanotube devices are relatively new, this situation may be improved on in the future, if higher conductivity nanotubes can be grown.
For realistic values of ${\mathcal R}$, the antenna efficiency is low. This is a drawback of nanotube antennas in the thin-wire geometry discussed in this paper. We return to this issue below.

\section{Impedance matching}

We wish now to discuss the input impedance and the issue of impedance matching.

\subsection{Numerical evaluation of input impedance}

Above, we found the antenna efficiency is maximized in the resonant case, which is therefore the most likely regime of operation. On this resonance, the imaginary part of the input impedance is {\it zero}. Therefore, one only has to deal with the real input impedance. We plot in Fig.~\ref{fig:rinvsreff} the input impedance, calculated from equation~\ref{eq:zinput}, as a function of ${\mathcal R}l$. We have again assumed that $l=0.01\lambda$.

In the low-loss case, the input impedance diverges, because the resonance causes there to be a null in the current at the antenna terminals. (See Fig.~\ref{fig:currentdist}) As the loss increases, the current at the terminals increases, thus decreasing the input impedance. When the input impedance becomes numerically equal to the characteristic impedance, this corresponds to the condition $\alpha l/2 \sim 1$. Above this value, the input impedance increases again, this time due to the severe ohmic losses. Interestingly, this curve is also a universal curve regardless of the frequency, the numerical value of $\mathcal{R}$, or the numerical value of $l$, as long as one assumes $l=0.01\lambda$. 

Off resonance, the imaginary impedance will be very large, and frequency dependent, making impedance matching more complicated.

\subsection{Natural transformer from free space to quantum impedance}

The input impedance is high, of order or larger than the resistance quantum of 25~k$\Omega$. This is no surprise, because in our model the antenna is fed with a high impedance two-nanotube transmission line. Thus, the nanotube antenna on resonance can be viewed as a {\it natural, quantum transformer}, that transforms the characteristic impedance of free space ($120\pi~\Omega$) up to the quantum impedance ($h/e^2$), assuming one is on resonance. Because most nano-devices and circuits are inherently high-impedance, this is a desired property of the system.

\section{Discussion}

\subsection{Loss and efficiency}

The fact that a nanotube current distribution has a different wavelength than the free space wavelength restricts its properties, in the best case, to be equivalent to those of a short thin-wire antenna. This means the radiation resistance is low, and causes any small ohmic resistance to reduce the overall antenna efficiency significantly. For current nanotube technology, this is a big challenge.

For impedance matching to quantum devices, it appears the optimum ${\mathcal R} l$ product is about 10~k$\Omega$, which translates into a resistance per length of about 10~$\Omega/\mu m$. This may be achievable in the near term at room temperature for nanotube antennas, and is likely achievable at cryogenic temperatures. According to our calculations, this would correspond to an antenna efficiency of about -60 dB. Clearly, this is not suitable for long-range wireless communications systems. However, it is generally a large unsolved problem to make electrical contact to the nanoworld, and even more difficult to transfer more abstract information from the macro-world to the nano-world. In this case, a wireless link from an integrated nano-system to the macroscopic world may still be advantageous over lithographic interconnects from a {\it systems} point of view, in spite of the somewhat low antenna efficiency. This problem of low-efficiency contact to the nanoworld is not unique to wireless interconnects. With dc contact to nano-devices the contact resistance is typically high and is a complicated physical phenomenon. From this perspective, wireless connections offer a much ``cleaner'' physical system, with less ambiguities such as Schottky barriers, quantum contact resistances, and similar issues currently being heavily investigated the field of nano-electronic devices.

\subsection{Transition from nano-antenna to thin-wire antenna}

A question which naturally arises in this context is: How thin does a wire have to be for its behavior as an antenna to be different than a regular thin-wire antenna? This can be re-phrased: at what diameter is the kinetic inductance comparable to the magnetic inductance? The magnetic inductance is only weakly dependent on the diameter, and is about 1~pH/$\mu$m. A rough estimate for how the kinetic inductance scales with length comes from observing that the kinetic inductance per unit length is crudely m/n$e^2$, where n is the number of electrons per length. If one assumes metallic systems with one electron per atom, and atoms of size 1 angstrom, then for a diameter of d, the number of electrons per meter along the wire is approximately $1/d^2$, with d in angstroms. Therefore, the kinetic inductance is approximately $(1/d^2)$~x~$10^{-11}H/m$. Equating this with the magnetic inductance and solving for d, we find that a value of $d=100~nm$ is the critical diameter demarcating the boundary between nano-antenna, where kinetic inductance dominates, and thin-wire antenna, where magnetic inductance dominates. There is clearly plenty of room to engineer antenna performance in the intermediate regime. This is an interesting topic for future study.

\subsection{Assumptions}

There are a number of unproven assumptions in this work. First, we argued that the effective circuit model of a two-nanotube transmission line includes kinetic inductance and quantum capacitance as dominant circuit elements. Second, we argued that the radiation does not affect the current distribution on the nanotube significantly. Third, we argued that the radiation {\it reactance} is small compared to the kinetic inductance, but did not explicitly calculate the radiation reactance. These arguments, while reasonable, should be put on more rigorous grounds through self-consistent calculations that include the full quantum properties of electrons in coupled two-nanotube systems and their interaction with microwave radiation. Our work should be viewed as an engineer's attempt to simplify a complicated physical system down to its most important basic elements in order to provide simplified approximations for antenna performance.

\subsection{Alternative geometries}

In our calculations, we have considered the simplest antenna geometry, that of a thin-wire antenna. In this case, the radiation resistance turns out to be very low, so that minimizing ohmic resistance is a critical issue. In other words, for resistive nanotubes, the antenna efficiency is low. However, our work is only the first step in the design of nano-antennas. For example, there may be other, alternative geometries that are more suited to particular properties of nanotubes, such as the high kinetic inductance and high resistance. This remains an open question for future work.

\subsection{Future work}

Our work is really only a baby-step in the field of the integration of wireless technology with nanotechnology. The next logical question is, {\it to what do you connect to the terminals of the antenna?} This is related to the nascent field of nano-electronics architecture, which has many issues remaining to be solved. In this context, our work in this paper provides initial steps in understanding the antenna properties of nanotubes and nanowires, which will be needed for the future architecture work. 

\section{Conclusion}

Simply speaking, one cannot think of a nanotube antenna 
in the same way as a thin-wire antenna because of the excess inductance of order $10^4$ time the inductance of a thin-wire antenna. This translates into performance predictions which are substantially different than thin-wire antennas, essentially because the wavelength of the current excitation is 100 times smaller than the wavelength of the far-field radiation, a unique situation.

An advantage of nanotube antennas is that the nanotube can serve as an excellent impedance matching circuit to get from free space to high impedance devices. A disadvantage, for current growth technology, is the low efficiency. With the nanotubes we are able to grow in our lab, we can achieve a predicted antenna efficiency of -90 dB. 

With future, higher mobility nanotubes, better performance would be possible, although prospects of approaching efficiencies of order unity seem dim with the simple thin-wire geometry considered in this work. (For this, alternative geometries may be required.)
Doing so will require nanotubes with ballistic transport over 100s of $\mu$m. That is maybe not totally unrealistic. After almost 30 years of research on MBE growth, it is now possible in 2DEGs at cryogenic temperatures to achieve ballistic transport over 100s of $\mu$m. The reduced phase space for scattering in CNTs makes it possible to have much higher mobilities than 2d systems, so it is conceivable to achieve. In that case, and for more realistic lossy cases, our theory provides quantitative predictions for expected nanowire and nanotube antenna performance.

\section*{Acknowledgment}
This work was supported by the Army Research Office (award DAAD19-02-1-0387), the Office of the Naval Research (award N00014-02-1-0456), DARPA (award N66001-03-1-8914), and the NSF (award ECS-0300557).

\section*{Appendix I}
\begin{figure}
\centering
\includegraphics[width=3.5in]{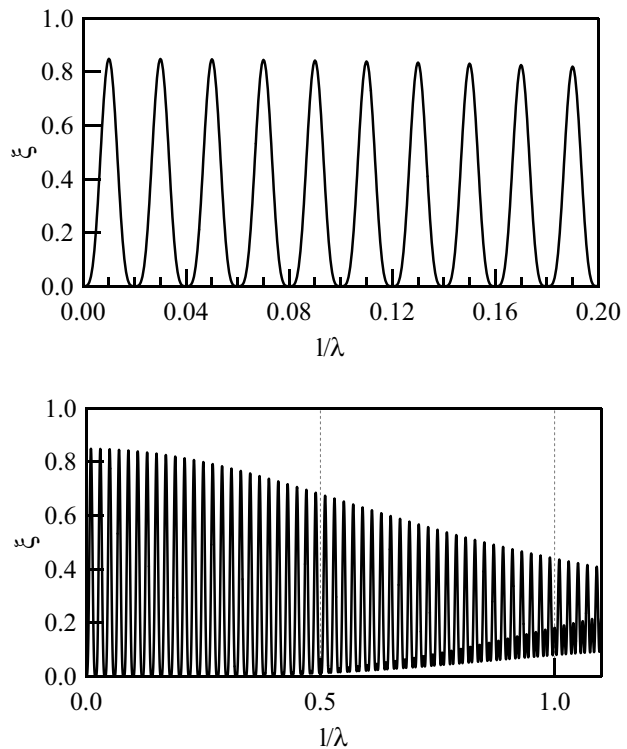}
\caption{Plot of $\xi$. The x-axis is $l/\lambda = kl/2\pi = 0.01 k_pl/2 \pi$.}
\label{fig:xi}
\end{figure}

In the Fig.~\ref{fig:xi}, we plot $\xi$ assuming $k_p=100 k$.

\section*{Appendix II}
\begin{figure}
\centering
\includegraphics[width=3.5in]{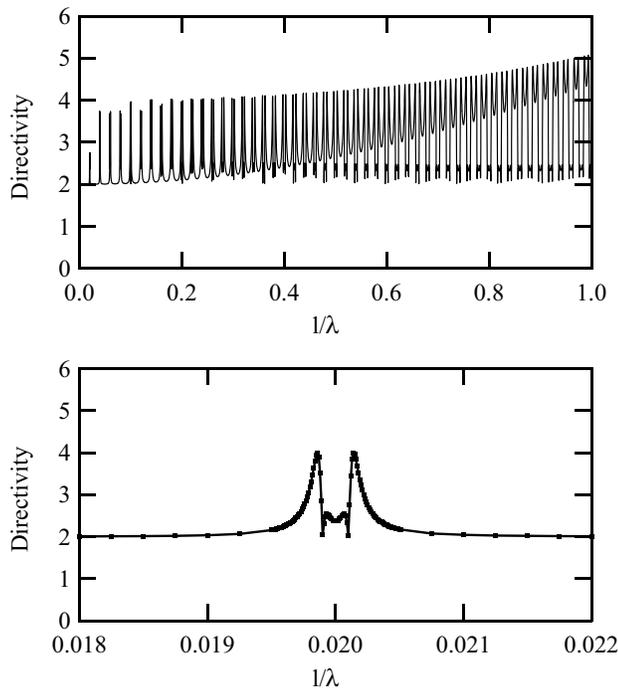}
\caption{Directivity. The x-axis is $l/\lambda = kl/2\pi = 0.01 k_pl/2 \pi$.}
\label{fig:directivity}
\end{figure}

In the Fig.~\ref{fig:directivity}, we plot the calculated directivity as a function of 
$l/\lambda = kl/2\pi = 0.01 k_pl/2 \pi$.

%\bibliography{nanoantenna}

%\begin{thebibliography}{1}

%H.~Kopka and P.~W. Daly, \emph{A Guide to {\LaTeX}}, 3rd~ed.\hskip 1em plus

%\end{thebibliography}

\begin{biography}[{\includegraphics[width=1in,height=1.25in,clip,keepaspectratio]{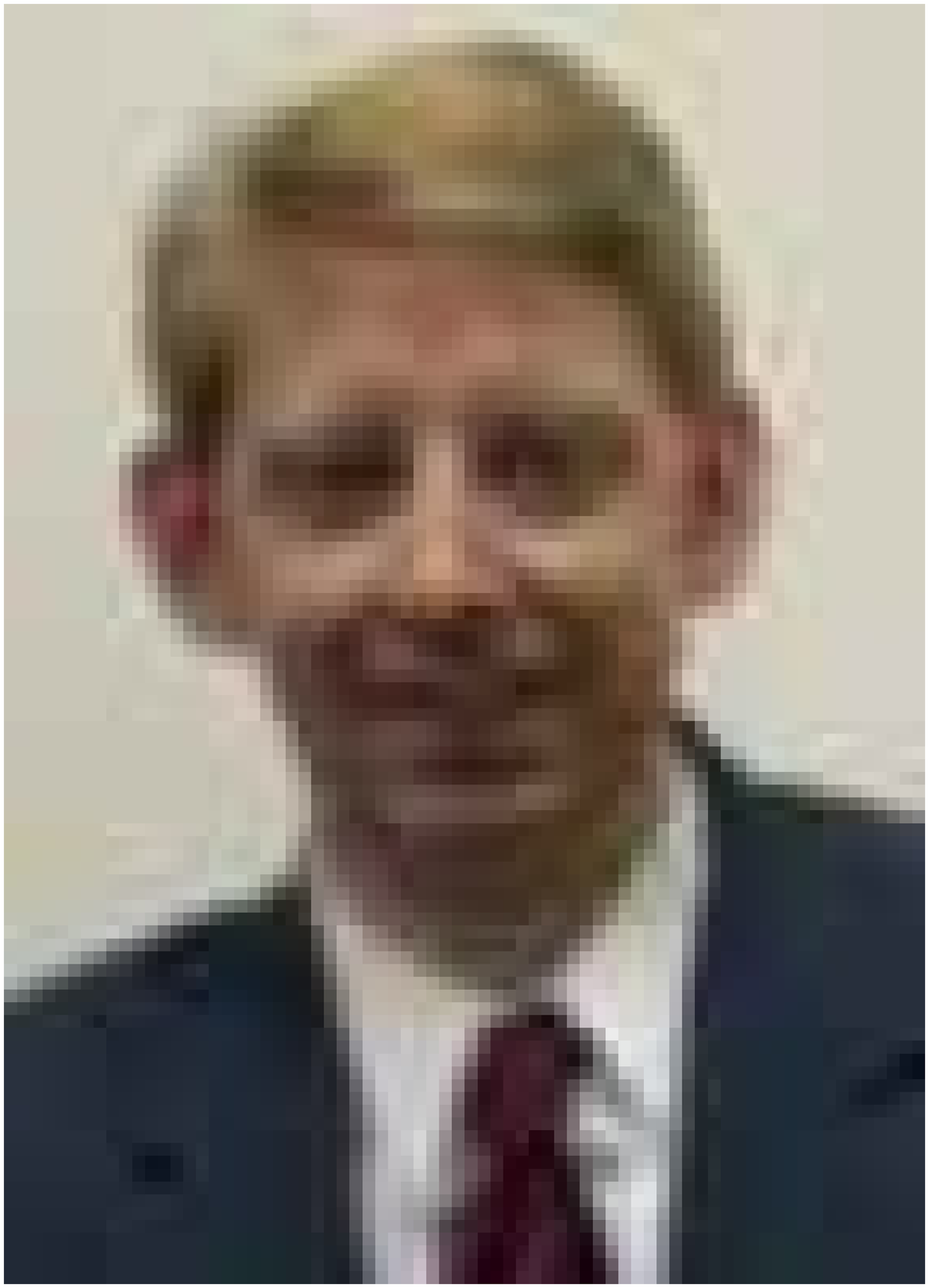}}]{Peter Burke}
received his Ph.D. in physics from Yale University in 1998.
From 1998-2001, he was a Sherman Fairchild Postdoctoral Scholar in
Physics at Caltech. Since 2001, he has been an assistant
professor in the Department of Electrical Engineering and Computer Science at
the University of California, Irvine.
\end{biography}

\begin{biography}[{\includegraphics[width=1in,height=1.25in,clip,keepaspectratio]{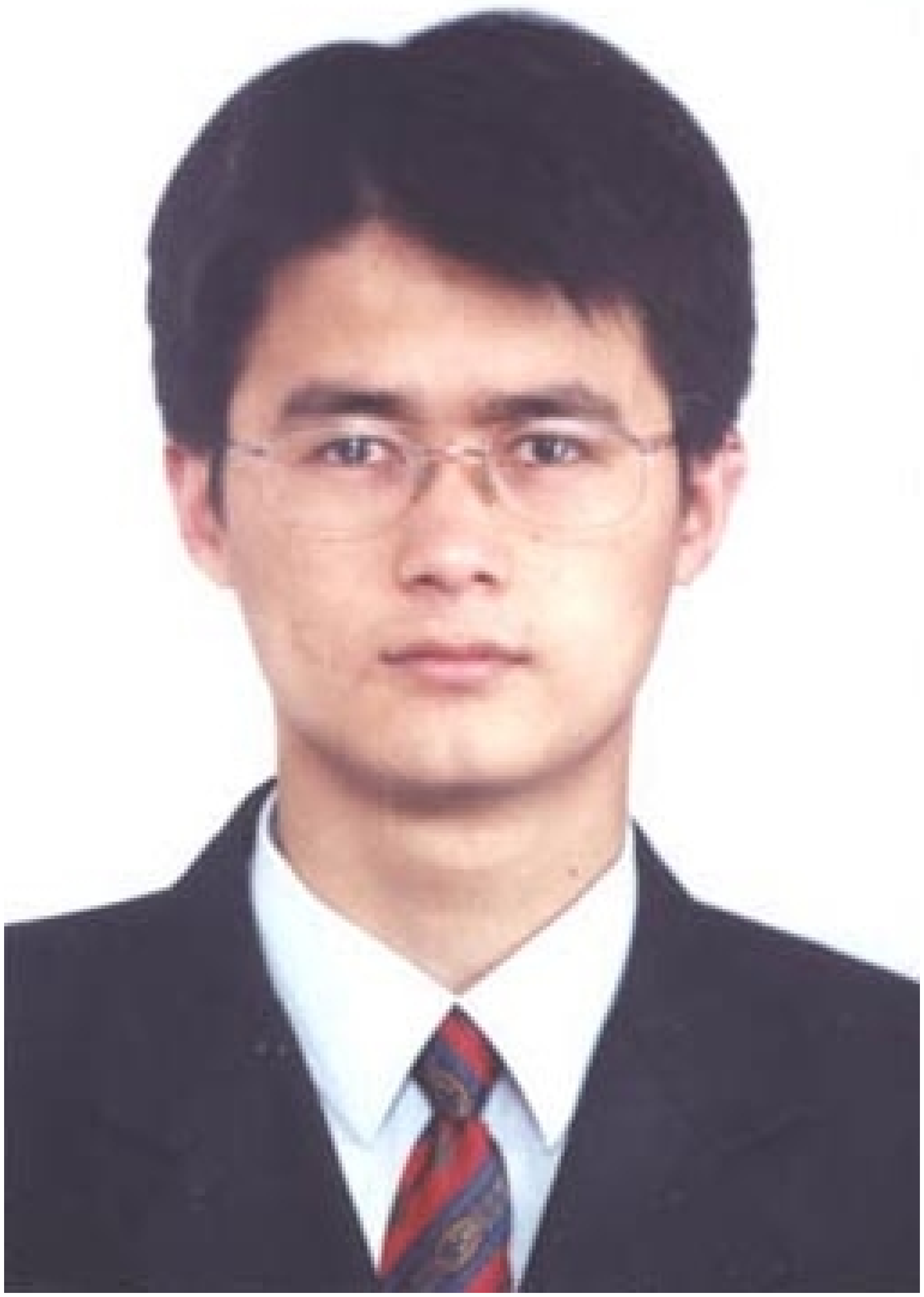}}]{Shengdong Li} received the M.S. degree in electrical engineering from the University of California at Irvine (UCI) in 2004. He is currently working toward the Ph.D degree in electrical engineering at UCI.
\end{biography}

\begin{biography}[{\includegraphics[width=1in,height=1.25in,clip,keepaspectratio]{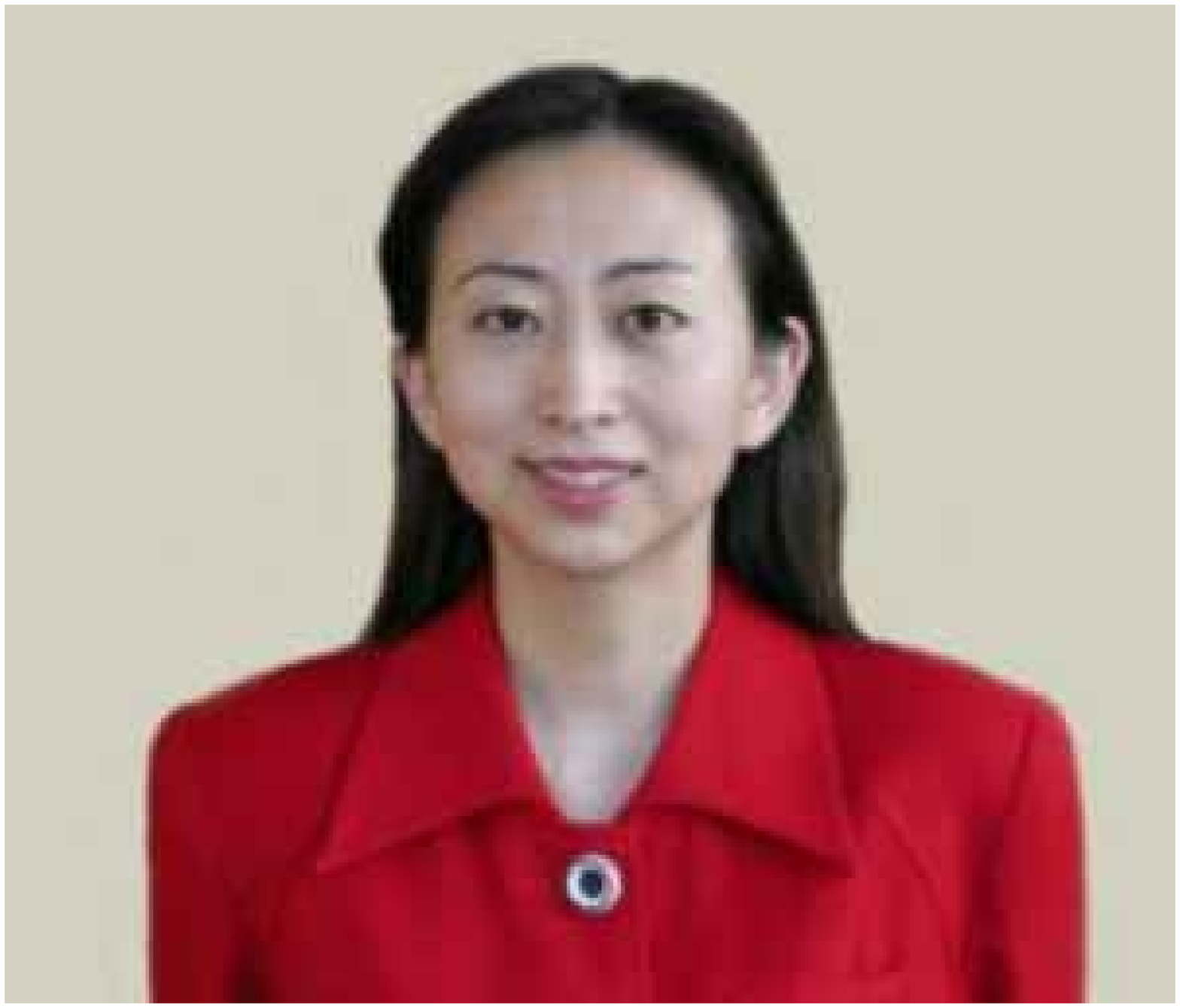}}]{Zhen Yu}
Zhen Yu has been a graduate student in EECS at U.C. Irvine since 2002.
\end{biography}

% you can push biographies down or up by placing
% a \vfill before or after them. the appropriate

%\vfill

% can be used to pull up biographies so that the bottom of the last one


\end{document}